\documentclass[]{aa}
\usepackage{graphicx}
\usepackage{natbib}
\usepackage{amssymb}
\usepackage{txfonts}

\bibpunct{(}{)}{;}{a}{}{,}
\newcommand{\be}{\begin{equation}}
\newcommand{\ee}{\end{equation}}
\newcommand{\phn}{\phantom{0}}

\begin{document}

\title{Microvariability in the optical polarization of
3C279\thanks{Based on observations made at the Complejo
Astron\'omico El Leoncito, which is operated under agreement
between CONICET and the National Universities of La Plata,
C\'ordoba, and San Juan, as well as at the Laborat\'orio Nacional
de Astrof\'\i sica, LNA-CNPq, Brazil.}\thanks{Table 2 is available
in electronic form at the  CDS via anonymous ftp to
cdsar.u-strasbg.fr (130.79.128.5) or via
http://cdsweb.u-strasbg.fr/cgi-bin/qcat?J/A+A/.}}

\author{I. Andruchow \inst{1}, S. A. Cellone \inst{2}, G. E. Romero
\inst{1,}\thanks{Member of CONICET}, T. P. Dominici \inst{3},
Z. Abraham \inst{3}}

\institute{Instituto Argentino de Radioastronom\'{\i}a, C.C.5, (1894) Villa
Elisa, Buenos Aires, Argentina
\and Facultad de Ciencias Astron\'omicas y Geof\'{\i}sicas UNLP, Paseo del
Bosque, B1900FWA La Plata, Argentina
\and Departamento de Astronomia, Instituto de Astronomia, Geof\'{\i}sica e
Ci\^encias Atmosf\'ericas, Universidade de S\~ao Paulo, Brasil}

\offprints{}
\date{Received / Accepted}

\titlerunning{Microvariability in the polarization of 3C279}
\authorrunning{I. Andruchow et al.}

\abstract{We present results of a microvariability polarization study
in the violently variable quasar 3C279. We have resolved the
polarization curves in the $V$ band for this object down to timescales
of minutes. We found two main components in the evolution of the
degree of linear polarization, one consisting of a flicker with
timescales of several tens of minutes and other component with far
more significant variations on timescales of a few days. The linear
polarization descended from $\sim 17$ \% down to $\sim 8$ \% in three
nights. The polarization angle underwent a sudden change of more that
10 degrees in a few hours, perhaps indicating the injection of a new
shock in the jet. The amplitude of the intranight flickering in the
degree of polarization is at the level of $\sim 1$\%. These are
probably the best sampled polarization data ever obtained for this
object. We also performed IR observations and we provide a follow-up
of the evolution of this source at such energies after the main
polarization outburst.  }

\maketitle

\section{Introduction}

The blazar 3C279 ($z=0.538$) is a strong source across the entire
electromagnetic spectrum and one of the best studied extragalactic
objects. Simultaneous multiwavelength observations from radio to gamma-rays
\citep[e.g.][]{Mar94,H96} show a spectral energy distribution dominated by
two non-thermal contributions interpreted as synchrotron and inverse Compton
radiation from relativistic electrons in the jet \citep[e.g.][]{H01a}. The
jet itself is well-resolved in multiepoch VLBI observations that allow to
determine its kinematics and the evolution of the superluminal components
\citep[e.g.][]{C93,P00,W01,P03}.

3C279 is extremely variable at all wavelengths. At optical bands it
has shown fast and significant outbursts in a single night
\citep[e.g.][]{MN96}. Rapid variations have been observed also at IR,
UV, X-ray, and gamma-ray energies by many authors \citep[e.g.][]{F99,
P99, K93, H01b}. Despite these studies, very little is known about the
short term polarization behaviour of this source. In this paper we
present, for the first time, optical polarization curves with high
temporal resolution for 3C279. We have found strong variations on
timescales of days superposed on a smaller amplitude flickering of
both the modulus and the angle of the $V$-band radiation polarization
vector. We also present new IR observations of this blazar that shed
light on the overall behaviour of the source on longer timescales.

The structure of the paper is as follows. In the next section we present a
description of the polarimetric observations and the data analysis. In
Section 3 we describe the main results on the polarization. Section 4
presents the infrared data. In Section 5 we discuss the origin of the
observed variability. Finally, we close with the conclusions in Section 6.

\section{Polarimetric observations and data analysis \label{s_obs}}

The observations were made using the CASPROF polarimeter on the 2.15-m
CASLEO telescope at El Leoncito, San Juan, Argentina during 4
consecutive nights, March 09--12, 2002. CASPROF was built at CASLEO,
based on the designs of the previous MINIPOL and VATPOL instruments
\citep[see][]{M84,Martinez}.  It is a rotating plate polarimeter with
a Wollaston prism that divides the incident light beam into two
components, each one directed to a different photomultiplier. The
instrument has an $UBVRI$-system filter wheel and a second wheel with
diaphragms of different apertures. The observations were carried out
using the Johnson $V$ filter and the $11.3$ arcsec aperture diaphragm.
Weather conditions were photometric, except for the second night
(March 10), when the observations were made through thin cirrus.

Standard stars were observed to determine the zero point for the
polarization angle and the instrumental polarization (although the
latter was found to be practically zero). These stars were chosen from
the catalog by \citet{T90}; their parameters are given in
Table~\ref{standars}.

\begin{table}[ht]
\caption{Standard stars \label{standars}}
\begin{tabular}{@{}lccccc}
\hline
\noalign{\vskip 2pt}
\hline
\noalign{\smallskip}
Name & $\alpha$(2000) & $\delta$(2000) & Type & $P_V$ & $\theta_V$\\
     & [{h:m:s}] & [{$^{\circ}$ $'$ $''$}] &  & [{\%}] & [{$^{\circ}$}]\\
\noalign{\smallskip}
\hline
\noalign{\smallskip}
HD64299 & 07:52:25.6 & $-23$:17:46 & zero & 0.151 & \ldots \\
HD98161 & 11:17:11.8 & $-38$:00:52 & zero & 0.017 & \ldots \\
HD298383 & 09:22:29.8 & $-52$:28:57 & angle & 5.23\phn & 148.6 \\
HD110984 & 12:46:44.9 & $-61$:11:12 & angle & 5.70\phn & \phn91.6 \\
\noalign{\smallskip}
\hline
\end{tabular}
\end{table}

\begin{table}%[t]
\label{mediciones} \caption{Results from polarization observations
of 3C279}
\begin{tabular}{lrcccrc}
\hline \noalign{\vskip 2pt} \hline \noalign{\smallskip} J.D. &
$P$~~ & $\sigma(P)$ & $\theta$ & $\sigma(\theta)$ &
$U/I$ & $Q/I$\\
2452342.0+  &   [$\%$]~  & [$\%$] & [$^{\circ}$]  & [$^{\circ}$] &
[$\%$] & [$\%$]\\
\noalign{\smallskip}
\hline
\noalign{\smallskip}
 0.68148 & 17.19 & 0.25 & 54.2 & 0.4 & 16.31 & $-$5.46 \\
 0.68850 & 17.38 & 0.19 & 54.9 & 0.3 & 16.36 & $-$5.88 \\
 0.69878 & 16.85 & 0.16 & 55.0 & 0.3 & 15.84 & $-$5.75 \\
 0.70824 & 16.84 & 0.17 & 54.0 & 0.3 & 16.02 & $-$5.20 \\
 0.71603 & 16.88 & 0.20 & 53.0 & 0.3 & 16.23 & $-$4.64 \\
 0.72365 & 16.85 & 0.13 & 52.5 & 0.2 & 16.27 & $-$4.37 \\
 0.73140 & 16.38 & 0.20 & 53.2 & 0.3 & 15.71 & $-$4.64 \\
 0.74005 & 16.00 & 0.14 & 53.3 & 0.3 & 15.33 & $-$4.58 \\
 0.74890 & 16.59 & 0.17 & 52.6 & 0.3 & 16.01 & $-$4.36 \\
 0.75778 & 16.83 & 0.15 & 52.9 & 0.3 & 16.19 & $-$4.60 \\
 0.76600 & 16.74 & 0.13 & 52.8 & 0.2 & 16.13 & $-$4.48 \\
 0.77330 & 16.44 & 0.14 & 52.0 & 0.2 & 15.95 & $-$3.95 \\
 0.78174 & 16.50 & 0.16 & 52.5 & 0.3 & 15.94 & $-$4.25 \\
 0.79057 & 16.64 & 0.14 & 53.1 & 0.2 & 15.99 & $-$4.63 \\
 0.79851 & 16.32 & 0.13 & 52.8 & 0.2 & 15.72 & $-$4.36 \\
 0.80610 & 15.58 & 0.20 & 52.0 & 0.4 & 15.12 & $-$3.76 \\
 0.81357 & 15.44 & 0.17 & 52.6 & 0.3 & 14.91 & $-$4.03 \\
 0.82140 & 16.28 & 0.17 & 52.8 & 0.3 & 15.68 & $-$4.37 \\
 0.82977 & 16.23 & 0.17 & 52.6 & 0.3 & 15.66 & $-$4.28 \\
 0.83895 & 15.44 & 0.14 & 53.0 & 0.3 & 14.84 & $-$4.26 \\
 0.84777 & 15.63 & 0.19 & 53.0 & 0.4 & 15.02 & $-$4.32 \\
 0.85538 & 16.00 & 0.17 & 52.8 & 0.3 & 15.41 & $-$4.31 \\
 0.86317 & 15.63 & 0.17 & 52.4 & 0.3 & 15.11 & $-$4.01 \\
 0.87163 & 15.08 & 0.16 & 52.4 & 0.3 & 14.58 & $-$3.84 \\
 0.87963 & 15.52 & 0.19 & 53.1 & 0.4 & 14.90 & $-$4.35 \\
\noalign{\medskip}
 1.45230 & 12.68 & 0.47 & 56.5 & 1.1 & 11.68 & $-$4.93 \\
 1.62768 & 13.87 & 0.22 & 55.5 & 0.5 & 12.95 & $-$4.96 \\
 1.63643 & 13.61 & 0.20 & 54.6 & 0.4 & 12.84 & $-$4.49 \\
 1.64633 & 13.60 & 0.22 & 54.1 & 0.5 & 12.92 & $-$4.23 \\
 1.65881 & 13.24 & 0.19 & 54.6 & 0.4 & 12.50 & $-$4.34 \\
 1.66966 & 13.15 & 0.20 & 55.1 & 0.4 & 12.34 & $-$4.55 \\
 1.67639 & 13.21 & 0.16 & 55.9 & 0.3 & 12.26 & $-$4.92 \\
 1.68305 & 13.75 & 0.20 & 55.8 & 0.4 & 12.79 & $-$5.06 \\
 1.69093 & 14.26 & 0.20 & 54.2 & 0.4 & 13.53 & $-$4.50 \\
 1.69875 & 14.14 & 0.17 & 54.0 & 0.4 & 13.45 & $-$4.37 \\
 1.70552 & 13.59 & 0.19 & 54.0 & 0.4 & 12.92 & $-$4.20 \\
 1.71250 & 13.58 & 0.16 & 53.3 & 0.3 & 13.01 & $-$3.97 \\
 1.74710 & 14.75 & 0.21 & 53.2 & 0.4 & 14.14 & $-$4.17 \\
 1.78440 & 14.68 & 0.43 & 54.1 & 0.8 & 13.94 & $-$4.61 \\
 1.79975 & 13.94 & 0.22 & 54.9 & 0.5 & 13.12 & $-$4.71 \\
 1.81310 & 14.16 & 0.20 & 55.4 & 0.4 & 13.24 & $-$5.02 \\
 1.82177 & 14.49 & 0.21 & 55.7 & 0.4 & 13.49 & $-$5.28 \\
\noalign{\medskip}
\hline
\end{tabular}
\end{table}

\setcounter{table}{1}
\begin{table}%[t]
\label{}
\caption{(cont.) Results from polarization observations of 3C279.}
\begin{tabular}{lrcccrc}
\hline \noalign{\vskip 2pt} \hline \noalign{\smallskip} J.D. &
$P$~~ & $\sigma(P)$ & $\theta$ & $\sigma(\theta)$ & $U/I$
& $Q/I$\\
2452342.0+&  [$\%$]~ & [$\%$] & [$^{\circ}$]  &  [$^{\circ}$] &
[$\%$] & [$\%$]\\
\noalign{\smallskip}
\hline
\noalign{\smallskip}
 2.62362 & 10.53 & 0.20 & 51.8 & 0.5 & 10.23 & $-$2.50 \\
 2.63138 & 10.21 & 0.20 & 52.0 & 0.6 &  9.91 & $-$2.45 \\
 2.63971 & 10.70 & 0.18 & 49.9 & 0.5 & 10.55 & $-$1.81 \\
 2.64744 & 10.47 & 0.20 & 50.1 & 0.5 & 10.30 & $-$1.86 \\
 2.65481 &  9.73 & 0.13 & 51.2 & 0.4 &  9.50 & $-$2.09 \\
 2.66241 &  9.91 & 0.12 & 51.4 & 0.4 &  9.66 & $-$2.19 \\
 2.66962 &  9.91 & 0.21 & 52.9 & 0.6 &  9.53 & $-$2.71 \\
 2.67718 & 10.18 & 0.13 & 52.3 & 0.4 &  9.85 & $-$2.58 \\
 2.68476 & 10.49 & 0.18 & 51.1 & 0.5 & 10.25 & $-$2.23 \\
 2.69229 &  9.76 & 0.17 & 50.4 & 0.5 &  9.59 & $-$1.84 \\
 2.69962 & 10.15 & 0.22 & 50.1 & 0.6 &  9.99 & $-$1.81 \\
 2.70675 & 10.66 & 0.18 & 50.8 & 0.5 & 10.44 & $-$2.13 \\
 2.71373 & 10.17 & 0.16 & 50.9 & 0.4 &  9.96 & $-$2.08 \\
 2.72169 & 10.21 & 0.16 & 51.1 & 0.4 &  9.98 & $-$2.16 \\
 2.73165 & 10.46 & 0.19 & 51.1 & 0.5 & 10.22 & $-$2.21 \\
 2.74151 & 10.27 & 0.17 & 52.1 & 0.5 &  9.95 & $-$2.53 \\
 2.75000 & 10.27 & 0.17 & 52.8 & 0.5 &  9.90 & $-$2.75 \\
 2.75840 & 10.49 & 0.20 & 52.0 & 0.5 & 10.18 & $-$2.53 \\
 2.76667 & 10.12 & 0.19 & 52.1 & 0.5 &  9.81 & $-$2.48 \\
 2.77430 &  9.91 & 0.16 & 51.8 & 0.5 &  9.64 & $-$2.32 \\
 2.78202 & 10.00 & 0.17 & 52.1 & 0.5 &  9.70 & $-$2.44 \\
 2.78978 &  9.97 & 0.13 & 52.6 & 0.4 &  9.61 & $-$2.63 \\
 2.79725 &  9.98 & 0.17 & 51.6 & 0.5 &  9.71 & $-$2.29 \\
 2.80483 &  9.59 & 0.14 & 52.7 & 0.4 &  9.25 & $-$2.55 \\
 2.81320 &  9.69 & 0.13 & 54.4 & 0.4 &  9.17 & $-$3.14 \\
 2.82248 & 10.11 & 0.15 & 53.8 & 0.4 &  9.64 & $-$3.06 \\
 2.83100 &  9.74 & 0.21 & 52.6 & 0.6 &  9.40 & $-$2.56 \\
 2.83924 & 10.20 & 0.22 & 51.6 & 0.6 &  9.93 & $-$2.34 \\
 2.84774 &  9.39 & 0.14 & 51.4 & 0.4 &  9.16 & $-$2.09 \\
 2.85558 &  8.83 & 0.14 & 50.9 & 0.5 &  8.64 & $-$1.80 \\
 2.86303 &  9.50 & 0.16 & 50.3 & 0.5 &  9.34 & $-$1.73 \\
 2.87104 &  9.73 & 0.17 & 50.8 & 0.5 &  9.53 & $-$1.96 \\
 2.87969 &  9.78 & 0.14 & 51.5 & 0.4 &  9.53 & $-$2.21 \\
\noalign{\medskip}

 3.61600 &  7.90 & 0.23 & 61.0 & 0.8 &  6.71 & $-$4.18 \\
 3.62431 &  8.78 & 0.16 & 60.3 & 0.5 &  7.55 & $-$4.47 \\
 3.63279 &  9.01 & 0.12 & 60.8 & 0.4 &  7.68 & $-$4.71 \\
 3.64130 &  9.13 & 0.19 & 61.3 & 0.6 &  7.70 & $-$4.92 \\
 3.64931 &  8.48 & 0.21 & 60.5 & 0.7 &  7.27 & $-$4.37 \\
 3.65723 &  8.13 & 0.18 & 62.0 & 0.2 &  6.74 & $-$4.56 \\
 3.66586 &  8.34 & 0.12 & 60.7 & 0.4 &  7.11 & $-$4.35 \\
 3.67504 &  8.84 & 0.10 & 59.5 & 0.3 &  7.73 & $-$4.29 \\
 3.68377 &  9.18 & 0.12 & 60.9 & 0.4 &  7.80 & $-$4.85 \\
 3.69201 &  9.18 & 0.14 & 60.9 & 0.4 &  7.80 & $-$4.84 \\
 3.70138 &  9.47 & 0.11 & 59.7 & 0.3 &  8.26 & $-$4.64 \\
 3.71044 &  9.61 & 0.13 & 60.3 & 0.4 &  8.28 & $-$4.88 \\
 3.71823 &  9.59 & 0.13 & 60.9 & 0.4 &  8.15 & $-$5.06 \\
 3.72931 &  9.39 & 0.15 & 60.4 & 0.4 &  8.06 & $-$4.82 \\
 3.74065 &  9.60 & 0.18 & 60.3 & 0.5 &  8.27 & $-$4.89 \\
 3.74885 &  9.95 & 0.19 & 58.8 & 0.6 &  8.81 & $-$4.62 \\
 3.75787 &  9.96 & 0.17 & 59.4 & 0.5 &  8.72 & $-$4.81 \\
 3.77062 &  9.72 & 0.14 & 61.4 & 0.4 &  8.16 & $-$5.27 \\
 3.78195 &  9.54 & 0.14 & 61.0 & 0.4 &  8.09 & $-$5.06 \\
 3.78927 &  9.68 & 0.15 & 60.8 & 0.5 &  8.24 & $-$5.08 \\
 3.79675 &  9.99 & 0.12 & 61.8 & 0.3 &  8.31 & $-$5.54 \\
 3.83752 & 10.45 & 0.15 & 61.7 & 0.4 &  8.71 & $-$5.77 \\
 3.84603 & 10.27 & 0.15 & 60.7 & 0.4 &  8.77 & $-$5.35 \\
 3.84695 &  9.87 & 0.14 & 59.4 & 0.4 &  8.66 & $-$4.74 \\
 3.85415 &  9.62 & 0.18 & 59.1 & 0.5 &  8.48 & $-$4.55 \\
 3.85585 & 10.03 & 0.15 & 60.0 & 0.4 &  8.68 & $-$5.03 \\
 3.86274 & 10.22 & 0.22 & 58.2 & 0.6 &  9.15 & $-$4.55 \\
 3.86730 & 10.06 & 0.24 & 57.5 & 0.7 &  9.12 & $-$4.25 \\
 3.87193 & 10.38 & 0.18 & 59.7 & 0.5 &  9.05 & $-$5.09 \\
 3.88070 & 10.40 & 0.17 & 60.0 & 0.5 &  9.01 & $-$5.20 \\
\noalign{\smallskip}
\hline
\end{tabular}
\end{table}

In Table 2
%(available only in electronic form),
we present polarimetric data obtained for 3C279 during our
campaign. Column (1) gives the Julian Date corresponding to each
observation. Columns (2) and (3) give the degree of polarization
and its associated error. Columns (4) and (5) present the position
angle and its error. Finally, in columns (6) and (7), we list the
normalized Stokes parameters.

\section{Variability results \label{s_mr}}

\begin{figure}
\includegraphics[width=\hsize]{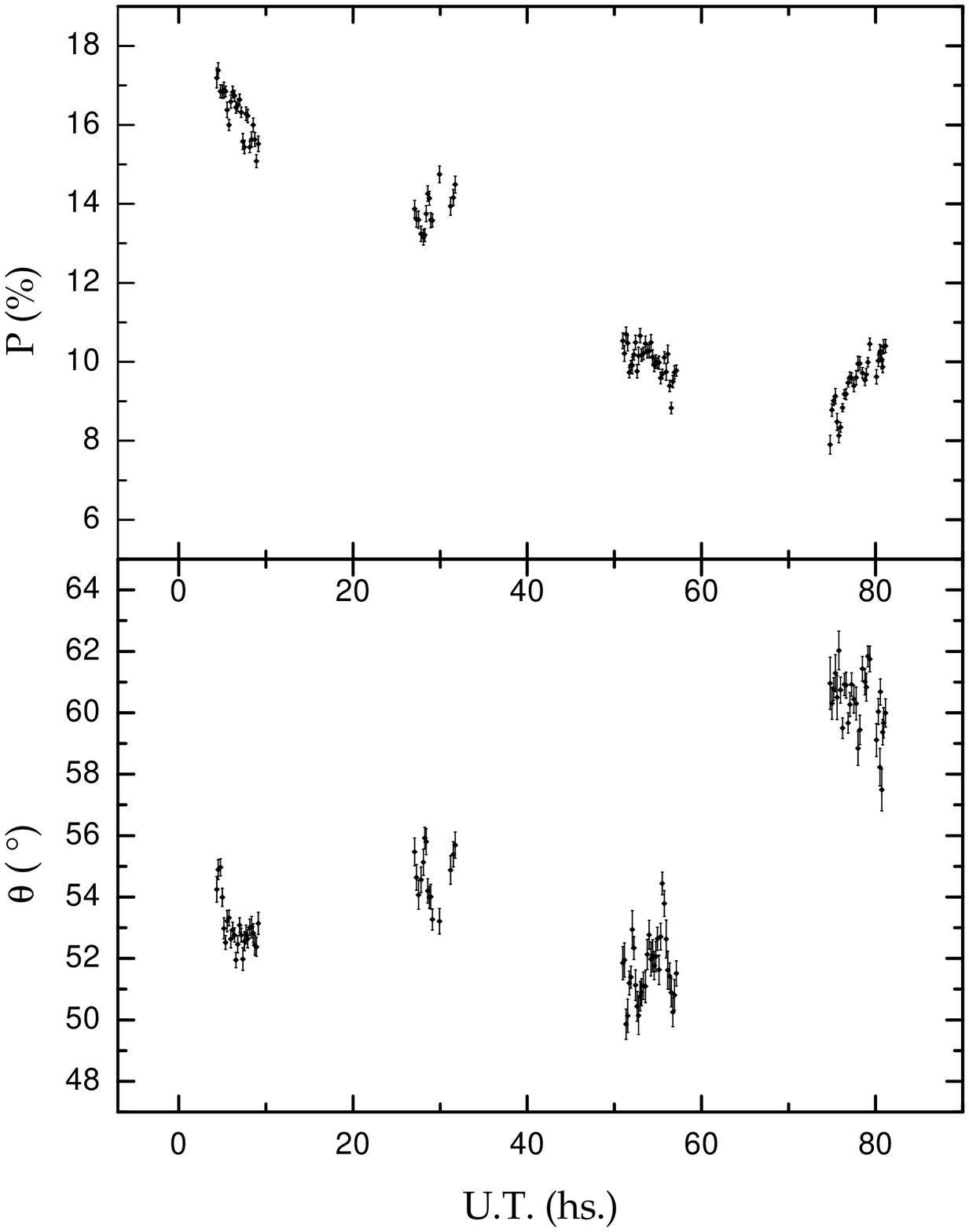}
\caption{Degree of polarization and position angle for the
observations of 3C279 during the four nights.} \label{campania}
\end{figure}

Figure~\ref{campania} presents the variation of the degree of
polarization (top panel) and position angle (bottom panel) during
the four nights of our campaign (time begins at $\mathrm{UT} = 0$
hs on March 09, 2002). After a first glance, we can see that there
are two different components in the variability with timescales of
hours and days, respectively. During the first three nights, the
degree of polarization drops from $\sim 17\%$ to $\sim 8\%$,
whereas in the last night it rises from $\sim 8\%$ to $\sim 10\%$.
With regard to the polarization angle, it also presents superposed
variations with different timescales. In this case, however,
during the first three nights the position angle shows no
large-amplitude changes, remaining between $51^\circ$ and
$53^\circ$, while in the fourth night there is a sudden increase
up to $\sim 63^\circ$.

\begin{figure*}
\includegraphics[width=0.45\hsize]{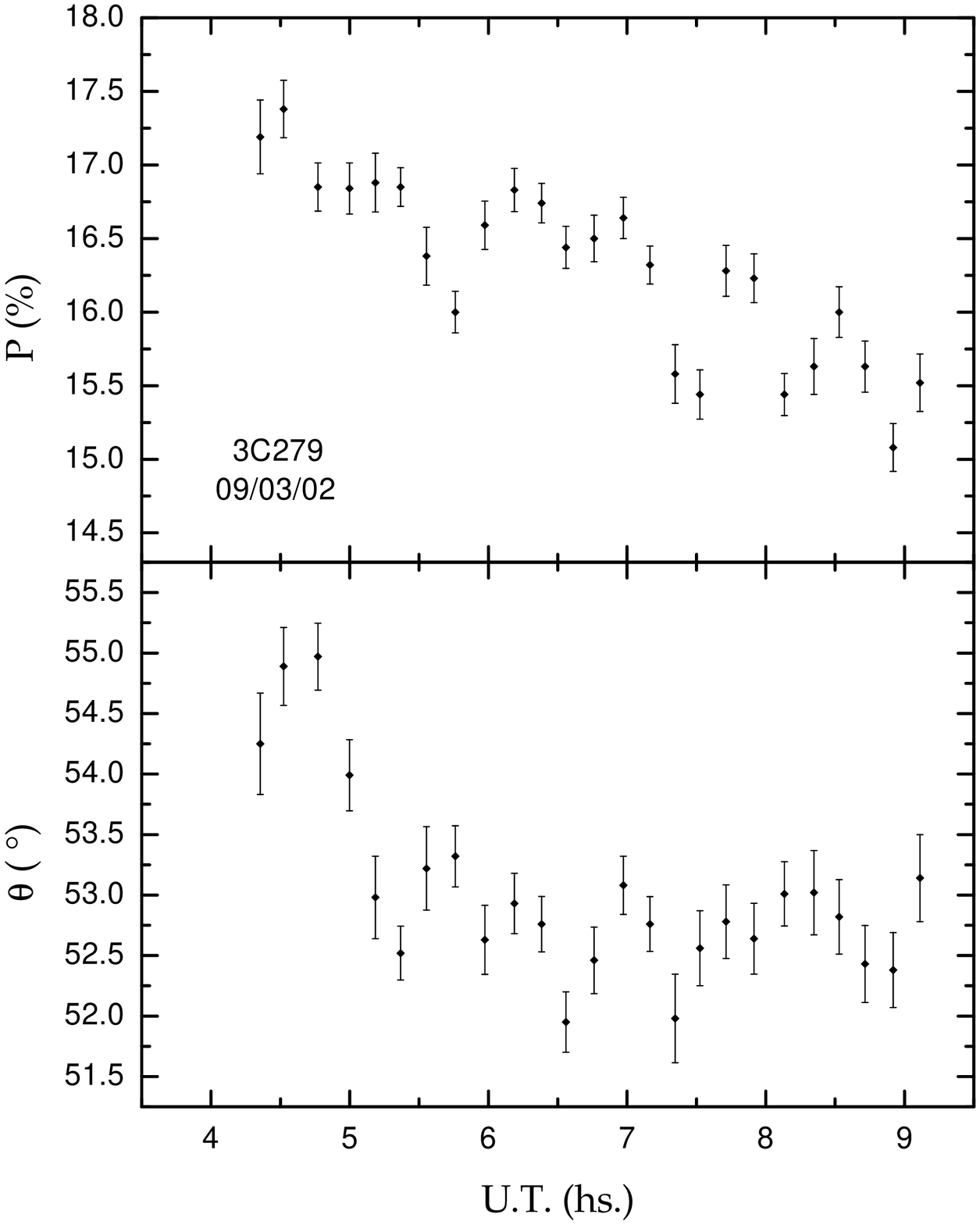}
\includegraphics[width=0.45\hsize]{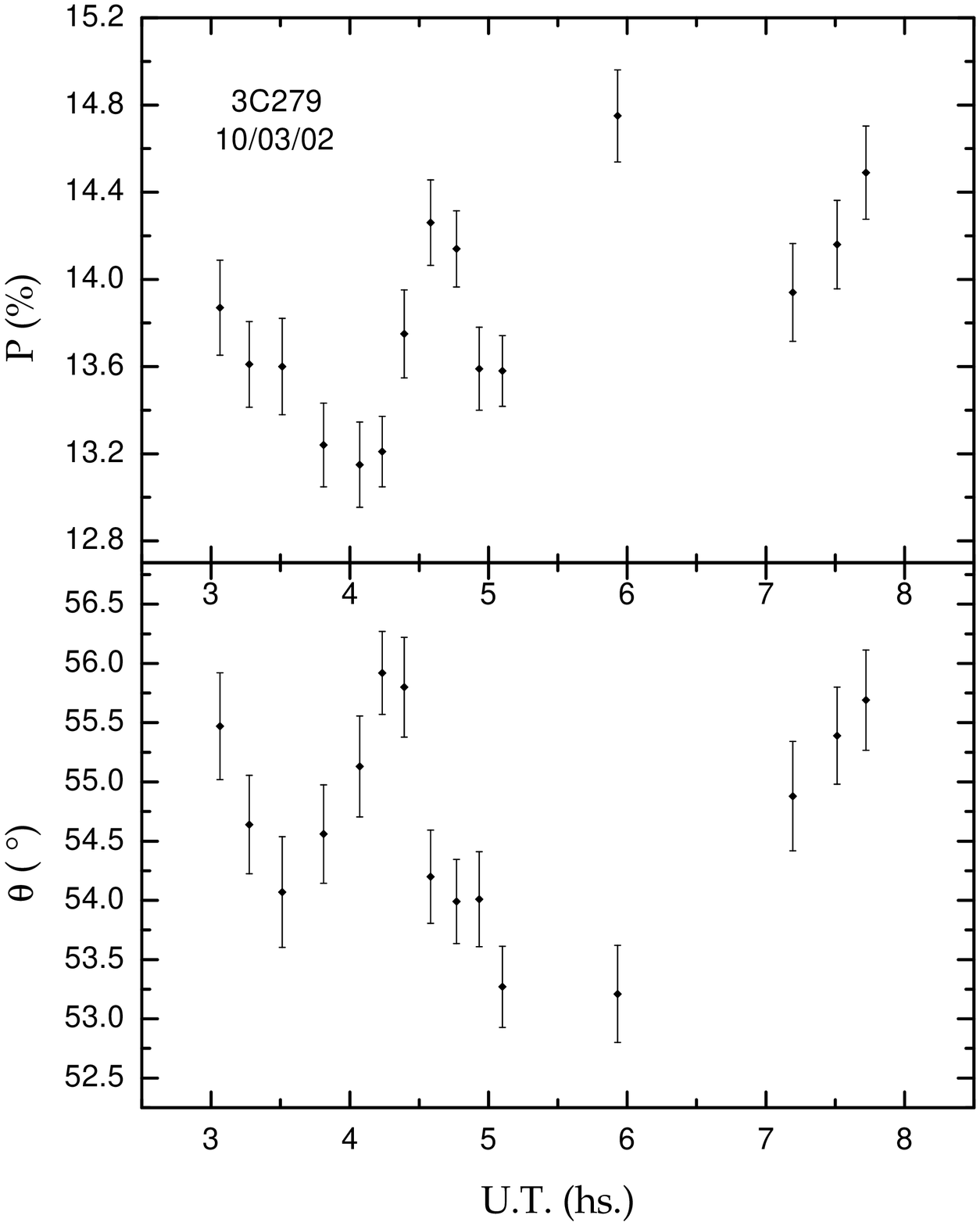}
\\[15pt]

\includegraphics[width=0.45\hsize]{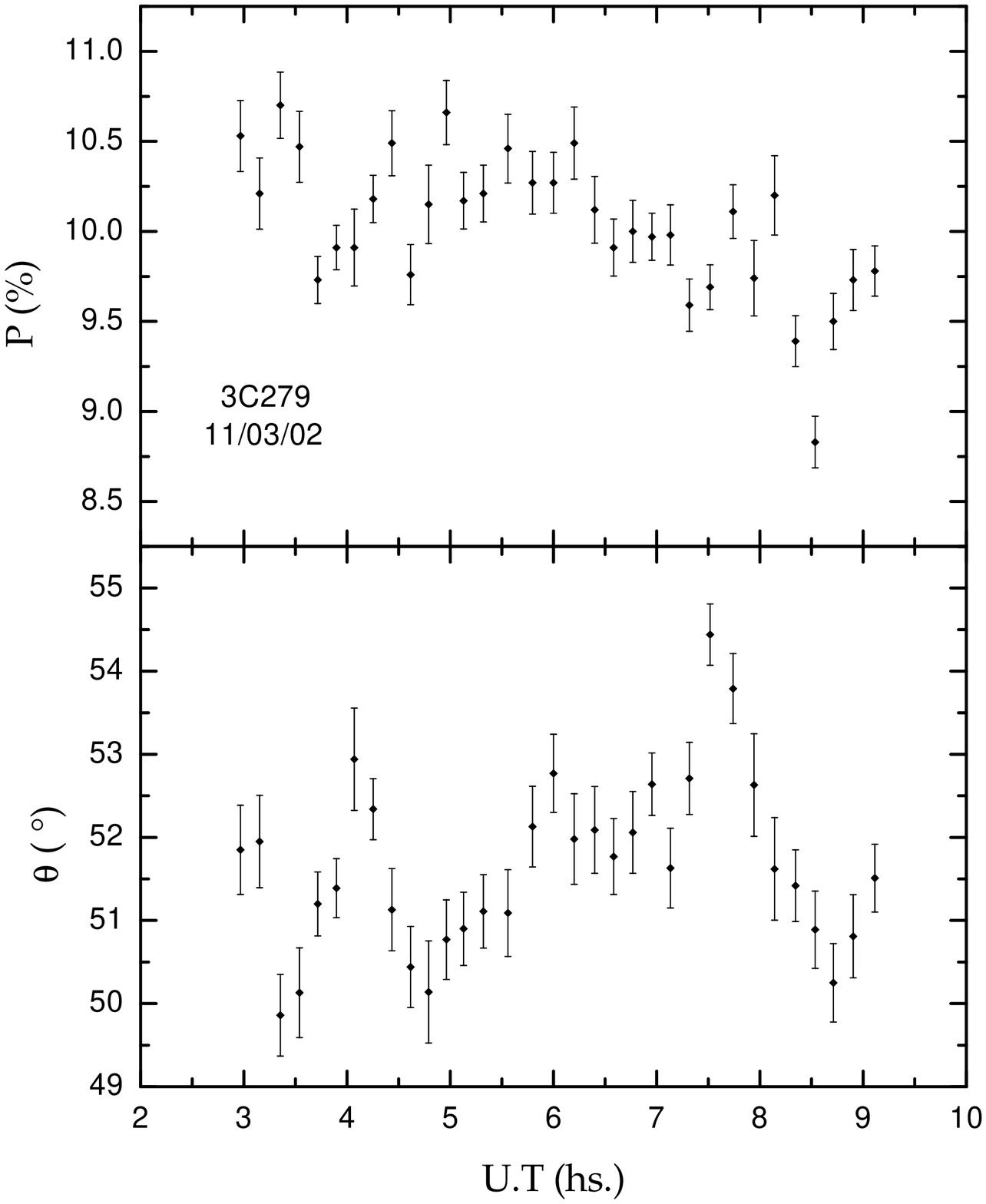}
\includegraphics[width=0.45\hsize]{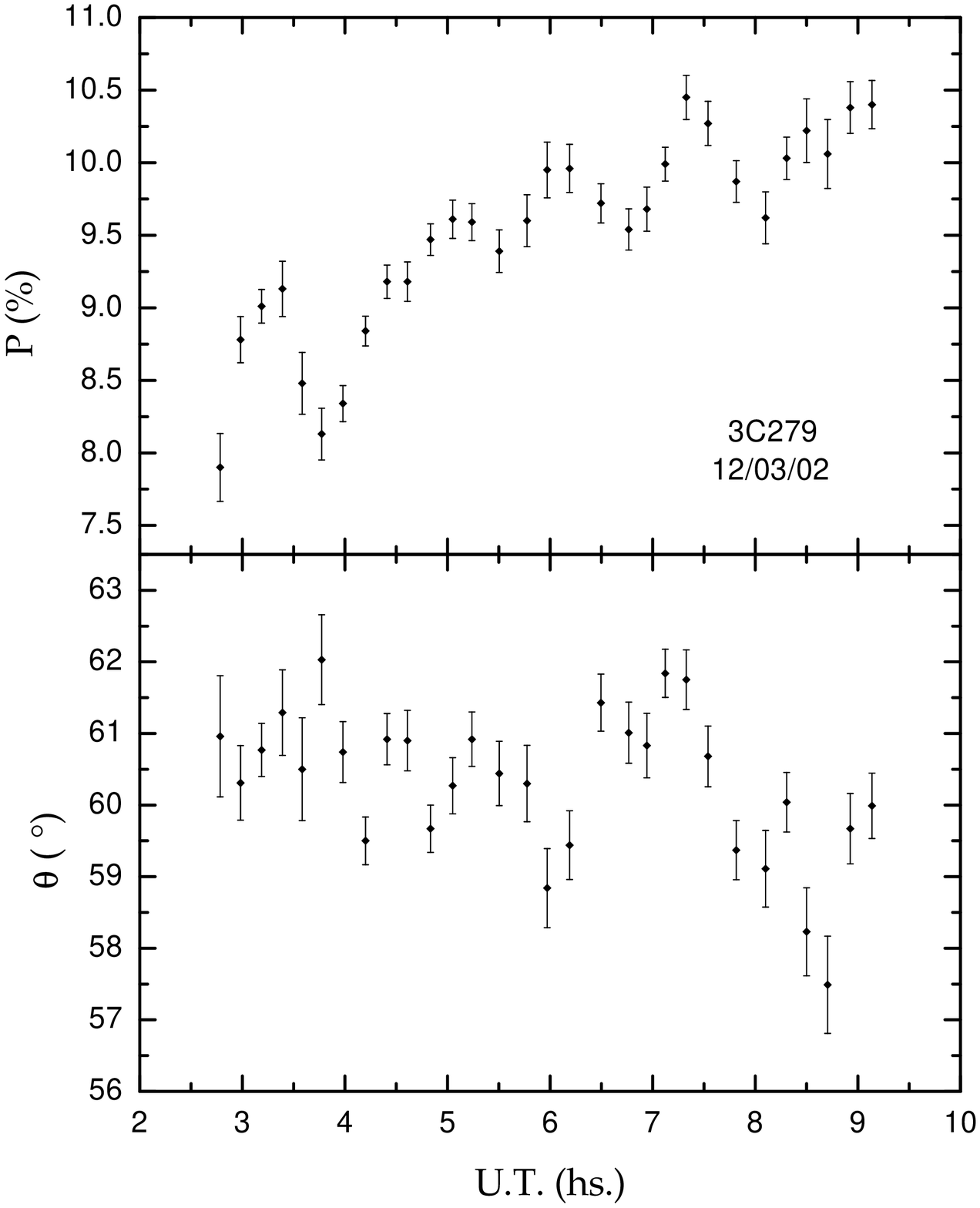}
\caption{\rm A detailed view of the polarization and position angle
variations for each night.} \label{each}
\end{figure*}

A detailed curve for each night is given in Figure~\ref{each} $a$ to
$d$, where the short timescale variations can be appreciated. These
can be described as a $\sim 1$ \% flickering in $P_V$ and variations
of up to $\sim 5$ degrees in $\theta_V$ within a few tens of minutes.

After discarding a few data points affected by moonlight
contamination and/or lightnings near the horizon during one night,
we made two different sets of data reductions. In a first approach
each data point corresponds to an individual integration. This method, however,
yields large error bars. As an alternative, we averaged each
pair of consecutive observations (in the $Q - U$ plane), in order to
improve the statistics. General trends were similar to those
obtained using the direct reduction, although the errors were
smaller. The rest of our analysis is thus based on these averaged data.

Since the polarimeter measures the ordinary and extraordinary rays almost
simultaneously, it is assumed that its results are immune to air
transparency and seeing changes. However, a possible error source lies in
the sky subtraction procedure.  To measure the polarization of the radiation
emitted by a source, two integrations are required: one with the object in
the center of the diaphragm, and the other of the sky near the source. Then,
the brightness and polarization vector for the sky are subtracted from the
corresponding source observation.

Large and/or rapid variations in the sky polarization due to the
presence of the Moon are expected. Since this factor affects both
the object and the sky, this systematic error should be removed
when the data are reduced, provided that each sky measurement is
made near (both in time and position) to the corresponding source
observation. The Moon was above the horizon only at the end of
each observing night; however, we checked for any residual
systematic error by plotting the behaviour of the polarization of
the object with respect to the polarization and magnitude of the
sky. No spurious variation seems to be present due to this effect,
neither for the polarization percentage nor for the position
angle.

In order to assess whether a source presents variability from a formal
and quantitative point of view, we followed the criterion of
\citet{K76} which has been used by several authors in variability
studies \citep{A82, R94}.  The variability, both in amplitude and
timescale, is quantified by the following parameters: the fluctuations
index $\mu$, the fractional variability index of the source $FV$, and
the time interval $\Delta t$ between the extrema in the polarization
curve.  The corresponding formulae are as follows:
\begin{equation} \mu = 100 \frac{\sigma_{\rm s}}{\langle
S \rangle}\, \;  {\%}, \end{equation}

\begin{equation}
FV = \frac{S_{\rm max} - S_{\rm min}}{S_{\rm max} + S_{\rm min}},
\end{equation}

\begin{equation}
\Delta t = | t_{\rm max} - t_{\rm min} |.
\end{equation}
Here $\sigma_s$ is the standard deviation of one observation, $\langle S
\rangle$ is the mean value of the polarization or the position angle
measured during the observing session, $S_{\rm max}$ and $S_{\rm min}$ are,
respectively, the maximum and minimum values for the polarization or the
position angle, $t_\mathrm{max}$ and $t_\mathrm{min}$ are the times when the
extreme points occur. Regarding the significance of the variability, a
source is classified as variable if the probability of exceeding the
observed value of

\begin{equation}
X^{2} = \sum_{i=1}^n{\epsilon_{i}^{-2}\, (S_i-\langle S
\rangle)^{2}}
\end{equation}
by chance is $<$ 0.1 \%, and non-variable if the probability is
$>$ 0.5 \%. If the errors are random, $X^2$ should be distributed
as $\chi^2$ whit $n-1$ degrees of freedom, where $n$ is the number
of points in the distribution.

In Tables 3 and 4 we show the values of the variability parameters
for the polarization percentage and the position angle,
respectively. Column 1 gives the Julian Date, Column 2, the number
of points for each night, Column 3 presents the value of $\mu$,
Column 4, $FV$, Column 5, $\Delta t$, Column 6 gives the mean
polarization (mean angle) for each night, Column 7, the value of
$\chi^2$, and Column 8 shows the variability class (V: if the
source is variable, NV: if it is not variable). The last row in
each table shows values for the whole four nights observing run.

\begin{table}
\caption{Variability results for 3C279: degree of
polarization}
\begin{tabular}{@{}lrccc@{~~~}c@{~~}c@{~}c@{~}}
\hline
\noalign{\vskip 2pt}
\hline \noalign{\smallskip}
~~J.D.& n\phn & $\mu$ & $FV$ & $\Delta${t} & $\langle P \rangle$ & $\chi^2$
&V/NV\\
      &   & [{\%}]&    & [{hs}] & [{\%}] & & \\
\noalign{\smallskip}
\hline
\noalign{\smallskip}
2452342 & 25 & \phn3.57 & 0.071 & 4.40 & 16.30 & \phantom{33}308.8 & V \\
2452343 & 17 & \phn3.54 & 0.075 & 3.08 & 13.78 & \phantom{333}91.0 & V \\
2452344 & 33 & \phn4.02 & 0.096 & 5.18 & \phn9.97 & \phantom{33}196.2 & V \\
2452345 & 30 & \phn6.48 & 0.139 & 4.54 & \phn9.47 &\phantom{33}503.4 & V \\
\noalign{\smallskip}
All & 103 & 72.77 & 0.375 & 70.26 & 11.68 & 33554.7 & V \\

\noalign{\smallskip} \hline
\end{tabular}
\label{Polari}
\end{table}

\begin{table}
\caption{Variability results for 3C279: position angle}
\begin{tabular}{@{}lrccc@{~~~}c@{~~}c@{~}c@{~}}
\hline
\noalign{\vskip 2pt}
\hline
\noalign{\smallskip}
~~J.D.& n\phn & $\mu$ & $FV$ & $\Delta${t} & $\langle \theta \rangle$ &
$\chi^2$ & V/NV \\
      & & [{\%}] &  & [{hs}] & [{$^{\circ}$}] &  &  \\
\noalign{\smallskip}
\hline
\noalign{\smallskip}
2452342& 25 & \phn1.38 & 0.028 & \phn1.79 & 53.0 & \phn157.0 & V \\
2452343& 17 & \phn1.69 & 0.030 & \phn3.08 & 54.7 & \phn\phn75.5 & V \\
2452344& 33 & \phn2.10 & 0.044 & \phn4.21 & 51.7 & \phn174.1 & V \\
2452345& 30 & \phn1.54 & 0.138 & \phn4.93 & 60.4 & \phn126.3 & V \\
\noalign{\smallskip}
All & 103 & 20.32 & 0.109 & 24.42 & 54.5 & 7721.9 & V \\
\noalign{\smallskip} \hline
\end{tabular}
\label{angulo}
\end{table}

\subsection{Stokes parameters}

An alternative way to present the results is through the
Stokes parameters. We shall consider, as usual, normalized
dimensionless parameters $U/I$ and $Q/I$.

\begin{figure}
\includegraphics[width=0.95\hsize]{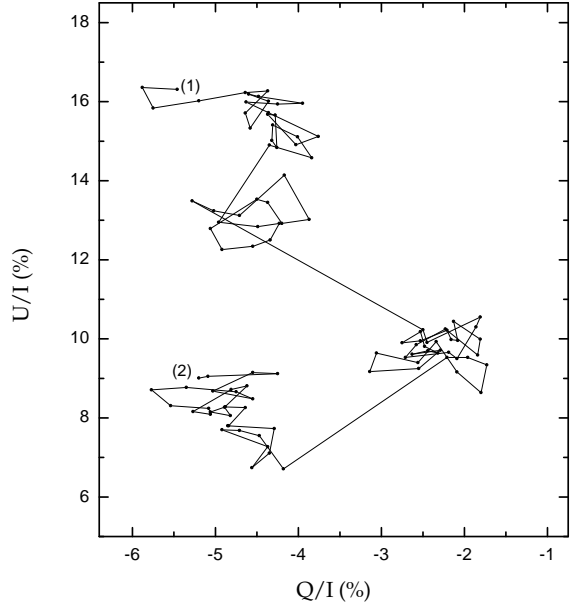}
\caption{Normalized Stokes parameters for 3C279 (entire campaign).
(1) indicates the initial data point, and (2) is the final one.}
\label{qvsutotal}
\end{figure}

\begin{figure*}
\includegraphics[width=0.45\hsize]{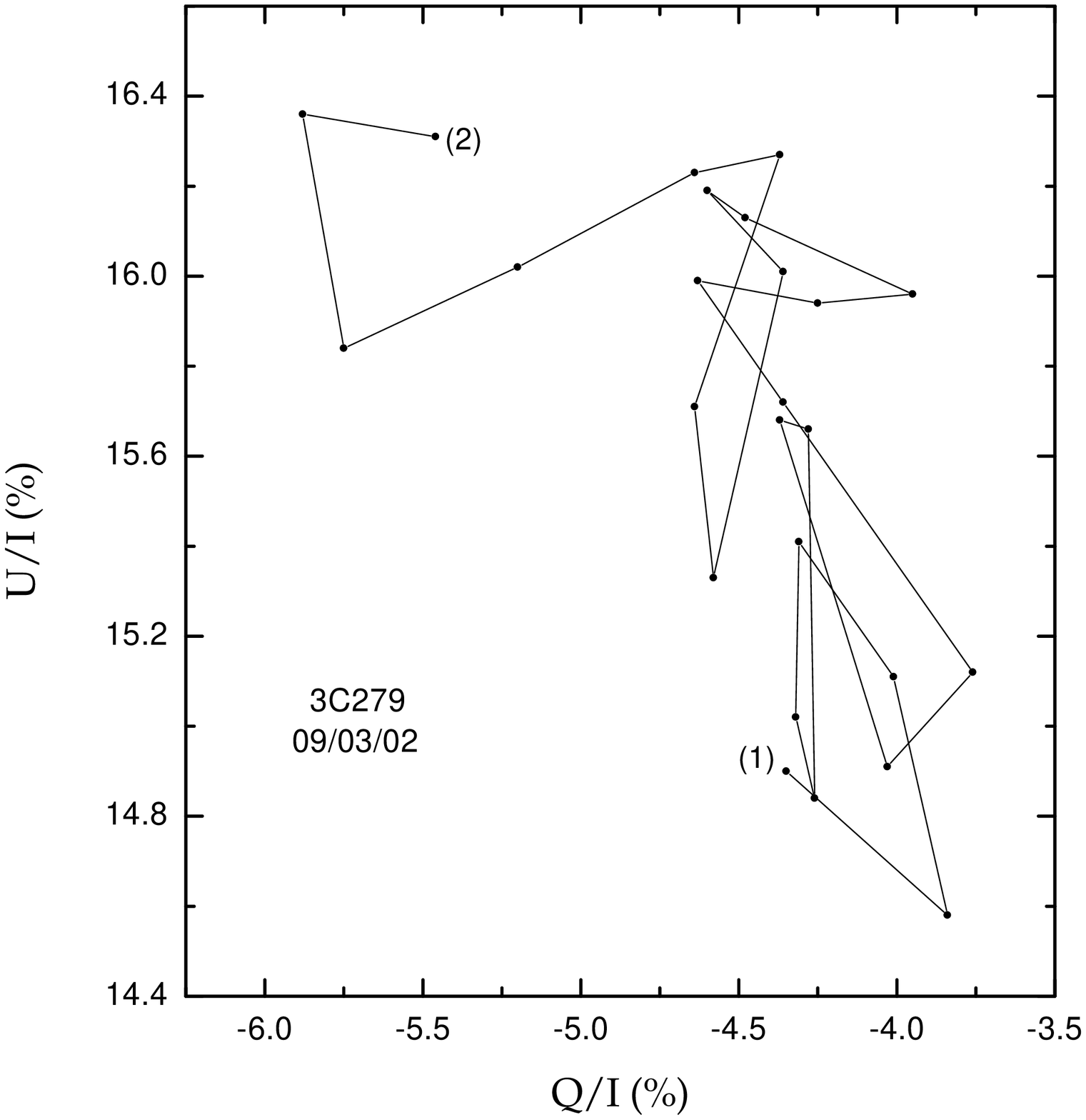}
\includegraphics[width=0.45\hsize]{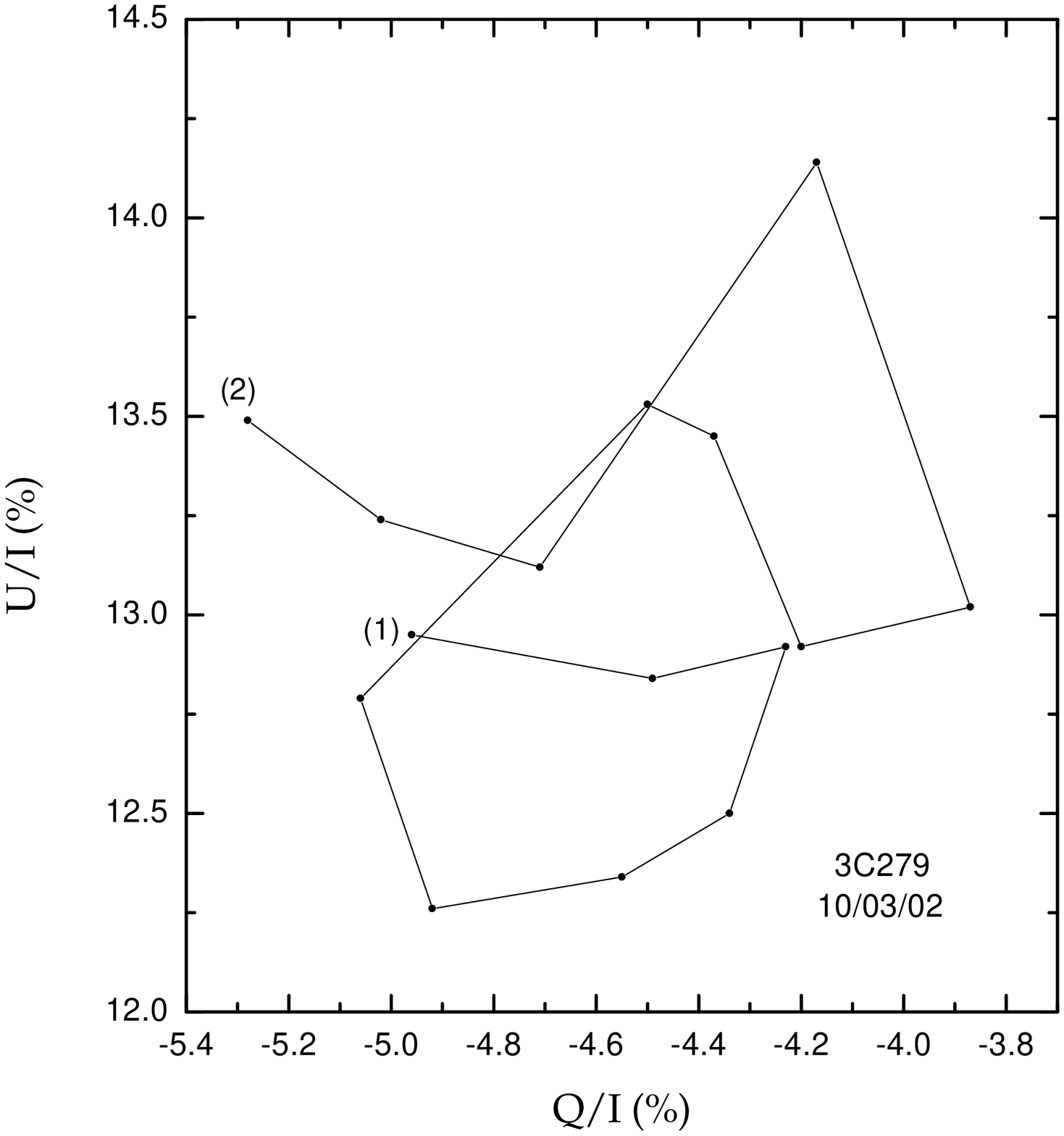}\\[45pt]

\includegraphics[width=0.45\hsize]{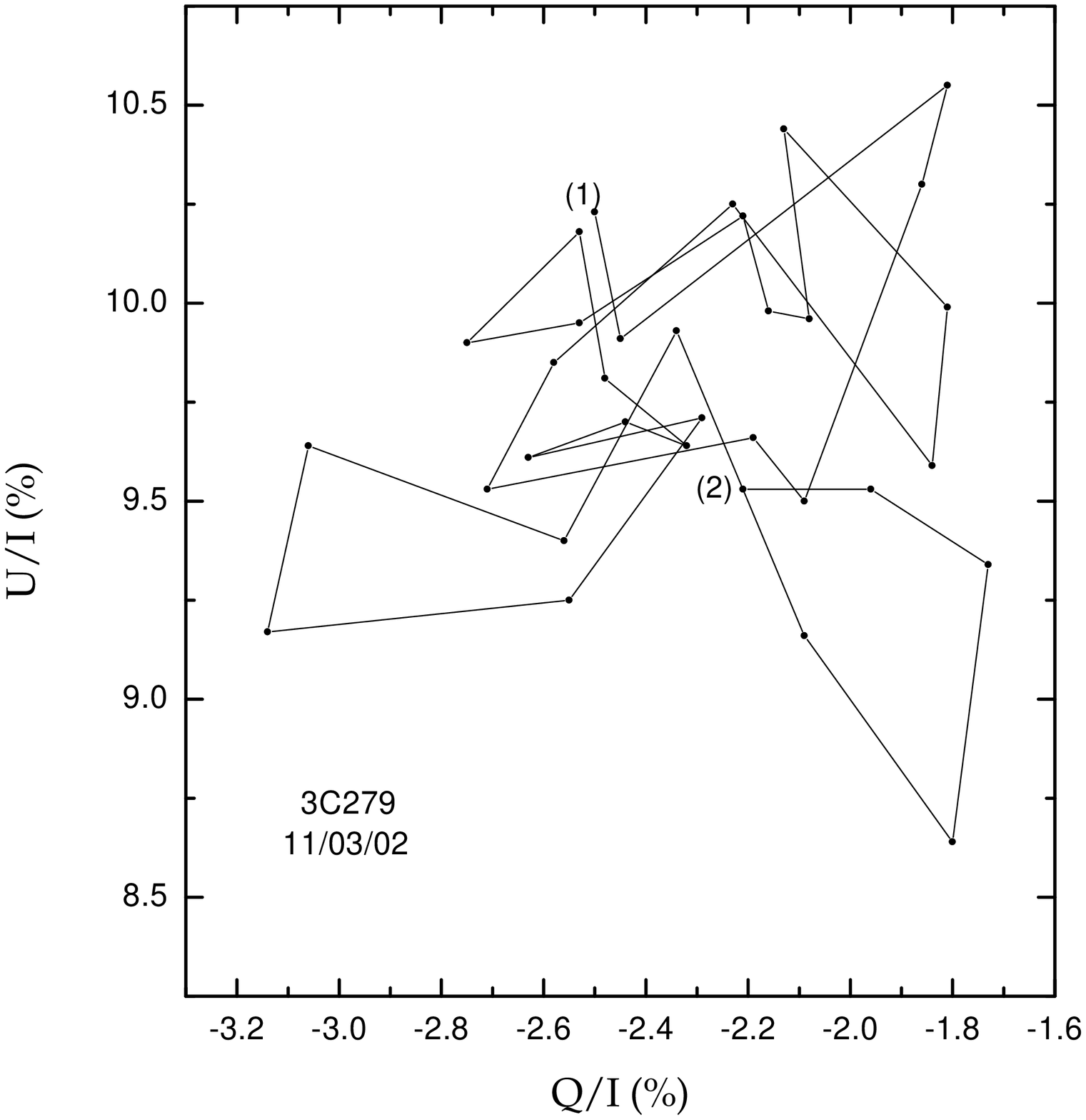}
\includegraphics[width=0.45\hsize]{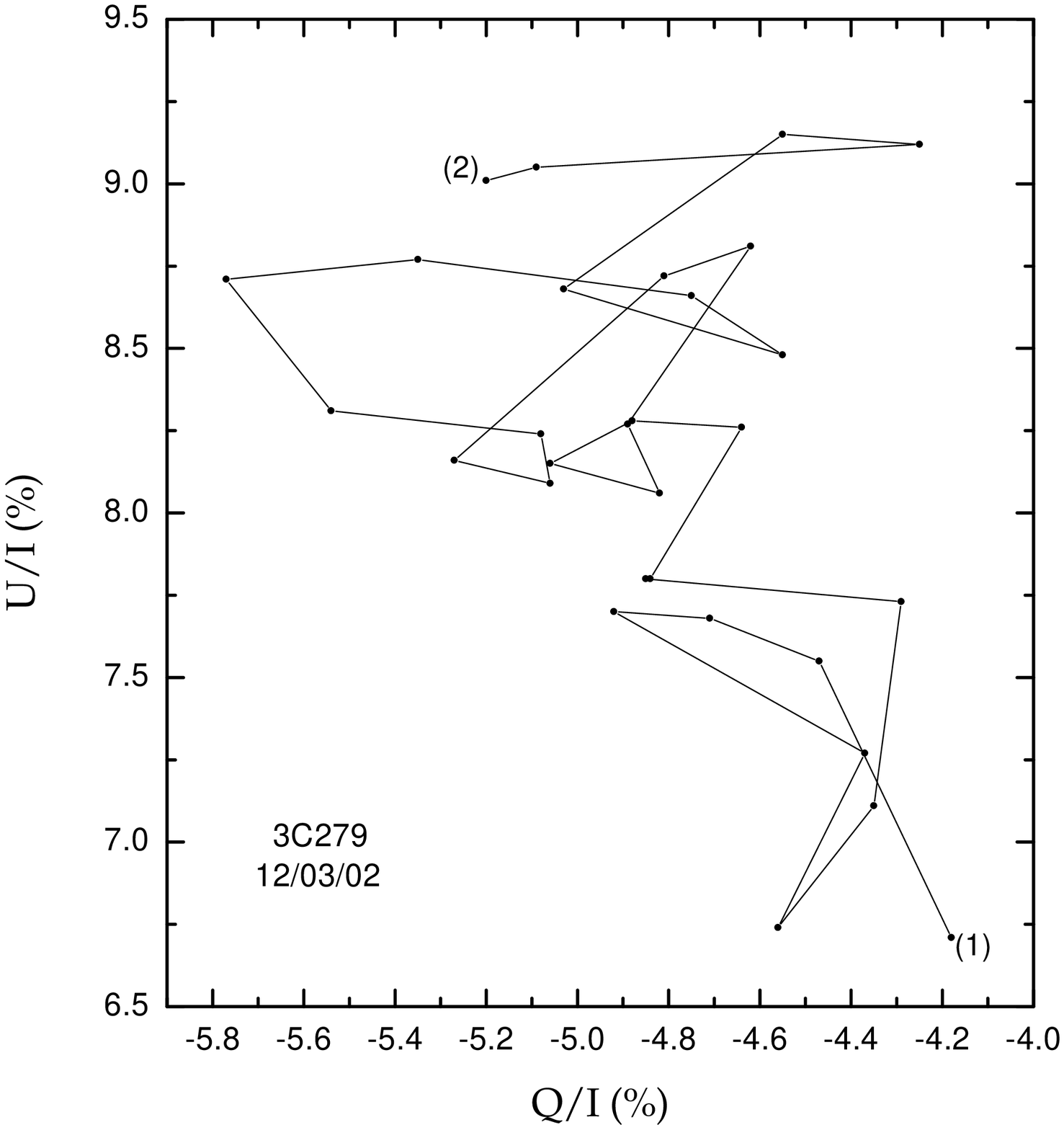}
\caption{\rm Stokes parameters for each night. Again, (1) and (2)
indicate the initial and final data point in each case.}
\label{qvsu}
\end{figure*}

In Figure \ref{qvsutotal}, we present the 3C279 data in the
$U/I$ -- $Q/I$ plane for the whole campaign. We can describe this pattern
like a random walk in this space. Figures~\ref{qvsu} $a$ to $d$ show
details for each night. In this expanded view we can also see that
there is no systematic pattern in the $Q-U$ plane, suggesting a
chaotic (turbulent) origin for the phenomenon.

\subsection{Visual Magnitude}

Additional information is provided by the visual magnitude of the source,
which was obtained simultaneously to each polarimetric measurement.
However, the intranight trend in the $V$ magnitude is within the expected
variations due to instrumental and atmospheric errors. In any case, no
strong correlation with the observed variations in polarization degree and
position angle is clearly evident.

On the other hand, there is a hint for an inter-night magnitude
variation.  We calculated the mean value of $m_{V}$ for each night and
for the whole campaign. These mean values for the photometric nights
are as follows: first night, $\langle m_{V}\rangle$ = 15.0; third
night, $\langle m_{V}\rangle$ = 15.1; last night, $\langle
m_{V}\rangle$ = 15.1; then the source dimmed by $~0.1$ mag. at the V
band during the four nights. Our mean magnitudes for 3C279 are thus
consistent with those obtained by \citet{S94} during an outburst
occurred in June 1992.

\section{Infrared observations}

The data in the near-infrared range ($J$ and $H$ filters) were taken with
the 1.6-m telescope at the LNA (Laborat\'orio Nacional de Astrof\'\i sica,
Braz\'opolis, Brazil), using the CamIV, an infrared camera with a
\textsc{hawaii} $1024 \times 1024$ pixel HgCdTe detector (18.5 $\mu$m
pixel$^{-1}$). The field of the resulting frames has 4 arcmin on each side,
with a scale of 0.24 arcsec pixel$^{-1}$. We obtained two frames of 120
seconds of total integration time each, with one arcmin offset between them,
to subtract the emission from the sky. Because of the poor weather
conditions, the observations were made only for a few hours in the nights of
March 9 and 11, 2002. Hence, a full lightcurve for the entire period
covered by the polarimetric campaign is not available, unfortunately.

\begin{figure}
\includegraphics[width=\hsize]{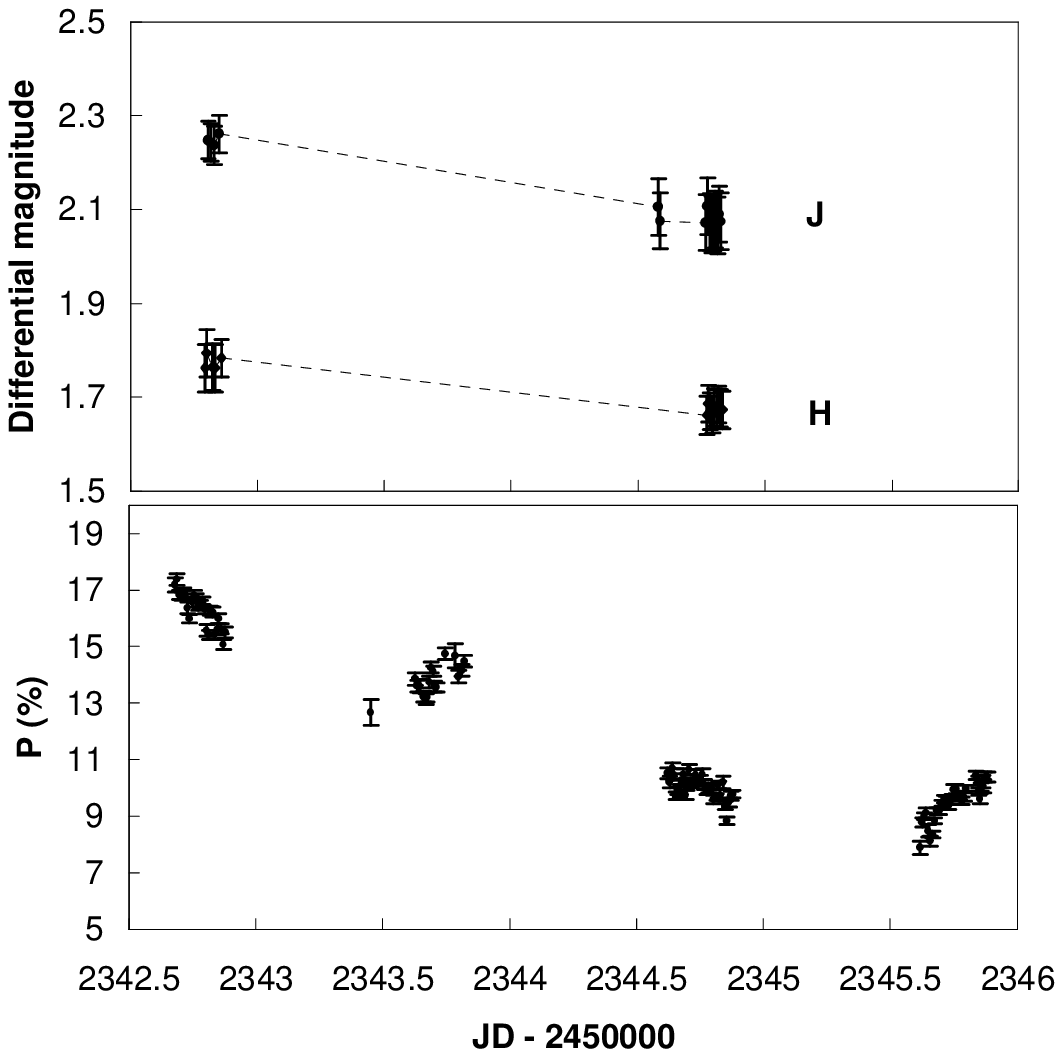}
\caption{\textbf{Upper panel:} Differential light curve of 3C279 in the $J$
and $H$ filters (comparison - target). \textbf{Lower panel:} Optical $V$ band polarization }
\label{ir}
\end{figure}

The reduction of the near infrared observations was made with the
\textsc{iraf}%
\footnote{\textsc{iraf} is distributed by the National Optical Astronomy
Observatories, which are operated by the Association of Universities for
Research in Astronomy, Inc., under cooperative agreement with the National
Science Foundation.}
tasks developed for the CamIV data analysis by F. Jablonski (private
communication, 2001). The two offset frames were combined after correcting
them by flat field, dark contribution and possible bad pixels in the
detector. The instrumental magnitudes were calculated using the
\textsc{apphot} task for aperture photometry, with an aperture of three or
four times the FWHM.

The lightcurves were constructed in differential mode, with
different field stars of similar magnitude to the object used for
control and comparison purposes \citep[see][for a detailed
discussion of the procedure]{Ce00}. The fluctuations in the lightcurves
that could be interpreted as intranight variations are within the
estimated errors. However, a brightness decrease between the two nights
could be observed, amounting to 0.13 mag in $J$ and 0.19 mag in
$H$. The differential lightcurves can be seen in Figure \ref{ir},
where the error bars correspond to the dispersion of the control
lightcurve. 3C279 is the only object in the field to show this
inter-night variation, which guarantees the confidence of the
detection. It is interesting to notice that the general trend of
the infrared variability mimics the behaviour of the degree of
polarization, with a qualitatively similar decrease.

\begin{figure}
\includegraphics[width=\hsize]{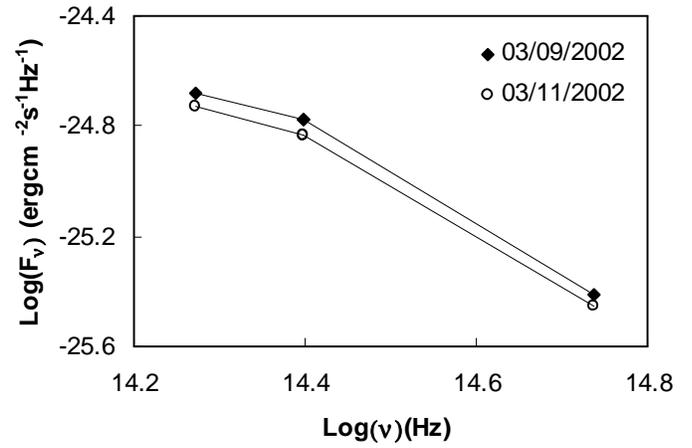}
\caption{Spectral behavior in the nights of March 9 and March
11, showing the flux decrease and that the mechanism that are
causing the variations (polarization and continuum flux) do not
cause changes in the spectral indices.} \label{irspec}
\end{figure}

The variability criterion adopted is the one used by \citet{J97},
\citet{RCC99}, and others: we calculated a parameter $C =
\sigma_{T}\diagup \sigma$, where $\sigma_{T}$ is the standard
deviation of the target differential lightcurve and $\sigma$ is
the standard deviation of the control lightcurve. A source can be
classified as variable with $99\%$ confidence level if $C\geq
2.576$. The analysis for data taken with both filters in the
infrared yields that the source formally classifies as
non-variable at all timescales except for averaged values from
night to night. Stars from \citet{Per98} were observed for flux
calibration and spectral index calculation. After correction for
redenning, we found that the near-infrared spectral index remained
constant in these two nights, as shown in Figure \ref{irspec},
with $\alpha_{ir} = -0.80 \pm 0.01$ (considering $F_{\nu} \propto
\nu^{\alpha}$).

\begin{figure}
\includegraphics[width=\hsize]{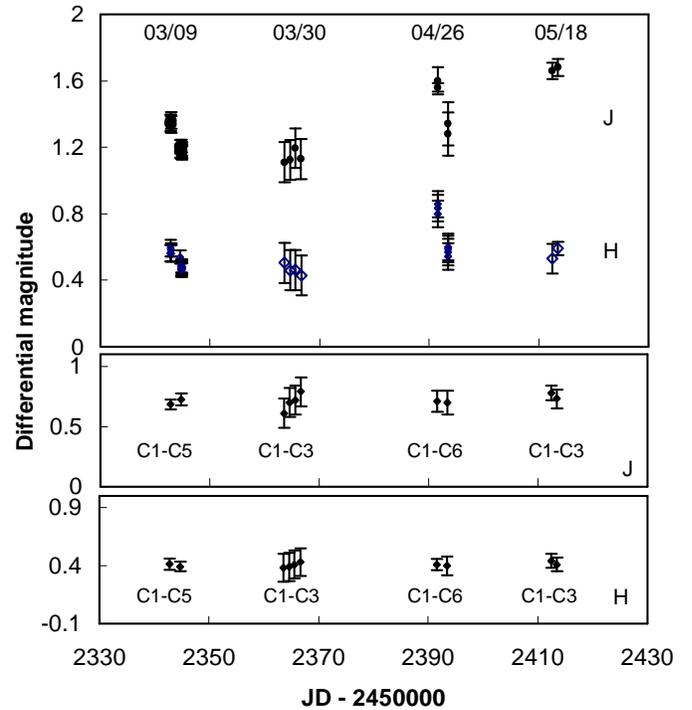}
\caption{Differential light curve of 3C279 in the $J$ and $H$
filters during ten nights of observations in 2002, divided in four
campaigns between March 9 and May 19.}
\label{irt}
\end{figure}

Since at the final of the coordinated campaign 3C279 showed an
increase in the degree of polarization as well as a sudden change
in the polarization angle, we decided to follow up its
near-infrared behavior in the subsequent months. The observations
were carried out with the CamIV installed in the 0.6-m B\&C
telescope at LNA (field of view of 8' x 8', with 0.47 arcsec
pixel$^{-1}$ resolution) for about one hour per night in three
campaigns during March 30 to April 2, April 26 and 29, and May
18-19. The resulting light curves in $J$ and $H$ bands can be seen
in Figure 7, which includes the first two nights (March, 9 and 11)
observation. The control light curves are also shown for the two
filters. The stars used for comparison were not the same for all
campaigns. As we already described above, in the first observing
run we used the 1.6 m telescope, with a small field of view. In
the subsequent campaign, the 0.6 m telescope was used and the
larger field allowed the use of a better sampled star. The star
chosen in campaigns 2 and 4 were not in the field in campaign 3,
and for that reason, a third comparison was used. In the figure,
the C1-C5 and C1-C6 control light curves were arbitrarily shifted
in relation to C1-C3 for better visualization. Finding charts with
the indications of the reference and control stars are available
upon request. The maximum brightness variation of 3C279 detected
in long time scales was about 0.4 magnitudes in the two filters,
between March 10 and April 29. In time scales of days, the object
infrared brightness decreased by 0.3 magnitudes between April 26
and 29.

In general, the behavior of $J$ and $H$ light curves was similar,
except in May, 2002, when the $J$ magnitudes increased in 0.3
magnitudes relative to its value is April 29, while the H magnitude
did not vary in the same period. This behavior was not observed in the
control light curves, which indicates that it must be real.

\begin{table*}
\caption{Results for the infrared observations of 3C279}
\begin{tabular}{cccccc}
\hline \noalign{\vskip 2pt} \hline \noalign{\smallskip} ~~Filter&
J.D. & $\sigma$ & $\Delta${t} & C & V/NV\\
     &  &   & [{hs}] &  \\
\noalign{\smallskip} \hline \noalign{\smallskip}
$J$ & 2452342 & 0.010 & 1.1 & 0.28 & NV \\
$J$ & 2452344 & 0.014 & 6.0 & 0.32 & NV \\
$H$ & 2452342 & 0.014 & 1.5 & 0.28 & NV \\
$H$ & 2452344 & 0.022 & 6.0 & 0.50 & NV \\
\noalign{\smallskip} \hline
\end{tabular}
\label{infra}
\end{table*}

\section{Discussion \label{s_disc}}

The polarized optical emission is expected to be synchrotron
radiation originated in the relativistic jet of 3C279. The rapid
variability, with timescales from minutes to hours, seems to favor
models based on the interaction of a relativistic shock with some
obstacles along the inner jet \citep[e.g.][]{G92}.

Small variations in the direction of the shocks that propagate down
the relativistic jet can produce large variations in the observed flux
and polarization. Recent evidence for changes in the trajectory of
superluminal components in 3C279 has been found by \citet{Ho03}
at kpc-scales. These changes can be quite frequent in the turbulent
environment of the inner pc-scale jet.

The fractional polarization of the shock as a function of the
angle with the line-of-sight affected by the relativistic
aberration, $\theta^{'}$, the compression factor of the plasma,
$k$, and the spectral index, $\alpha$, is \citep[e.g.]{Hu85}:

\begin{equation}
\Pi =
\frac{1-\alpha}{5/3-\alpha}\;\frac{(1-k^{2})\sin^{2}\theta^{'}}{2-(1-k^{2})\sin^{2}\theta^{'}}.
\end{equation}

The relation between $\theta^{'}$ and the actual viewing angle,
$\theta$, is given by:

\begin{equation}
\cos  \theta^{'} = \frac{\cos \theta - \beta}{1 - \beta \cos
\theta},
\end{equation}
where $\beta = (1 - \Gamma^{-2})^{1/2}$ is the relativistic velocity
of the shock and $\Gamma$ is the Lorentz factor of the shocked plasma.

In the case of a perpendicular,
relativistic strong shock, the factor by which the jet plasma is
compressed by the shock, $k$, can be written as \citep{B76}:
\begin{equation}
k^{-1} = \frac{n_{\rm ps}}{n_{\rm j}}=\frac{\hat \gamma  \Gamma_{\rm ps} + 1}{\hat \gamma - 1}.
\end{equation}
In this expression, $n_{\rm ps}$ and $n_{\rm j}$ are the particle
densities in the shocked gas and in the underlying jet, respectively,
$\hat \gamma$ is the adiabatic index, which has a value $\hat \gamma=
13/9$ for a gas with equal fractions of relativistic electrons and
non-relativistic protons, and $\Gamma_{\rm ps}$ is the Lorentz factor
of the shocked gas measured in the frame of the unshocked gas. If we
take the same velocity for the shock and the shocked plasma in order
to simplify the equations, $\Gamma_{\rm ps}$ can be expressed as
\citep[see][]{R95}:

\begin{equation}
\Gamma_{\rm ps} \simeq \frac{\Gamma \gamma_{\rm j} (1 -\beta^{2})}{\sqrt{2}}.
\end{equation}
Here $\gamma_{\rm j}$ is the Lorentz factor of the underlying jet.
Since the superluminal components are expected to be extremely
relativistic even respect to the jet, we adopt a mild value $\gamma_{\rm
j}\sim 2$ for the stable, underlying flow \citep{M85, R95}.

In the particular case of 3C279, we assume an average value of
$\alpha=-0.95$ for the spectral index at optical wavelengths \citep{D83}.
Concerning the Lorentz factor, we adopt as an average value the result of
the fit of superluminal components with the precessing jet model proposed
by \citet{A98}. This model yields $\Gamma = 13$ for $H_0 = 70$ km s$^{-1}$
Mpc$^{-1}$. We consider values of $\Gamma$ of $8$, $12$, $13$ and $16$.
The aberrated viewing angle at the present epoch is $\theta^{'}\sim 26\deg$,
which is in good agreement with the value given by \citet{P03},
once the aberration is corrected (i.e. $\theta\sim 2\deg$).

\begin{figure}
\vspace{2cm}
\includegraphics[width=0.8\hsize]{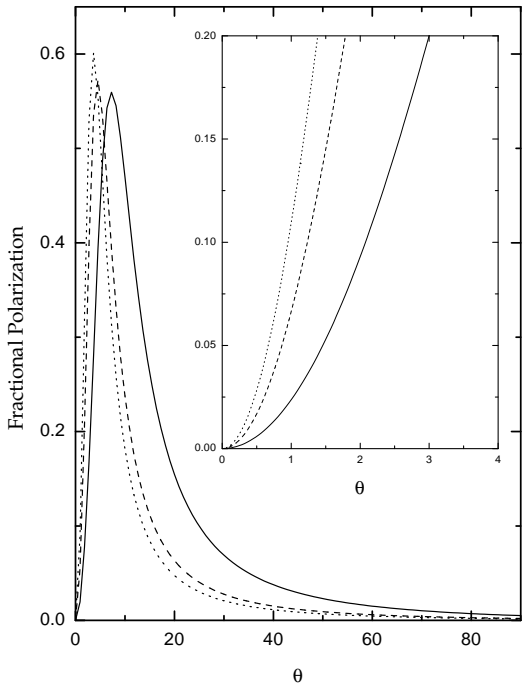}
\caption{Variation of the fractional polarization as a function of
the angle with the line-of-sight. The solid line is for $\Gamma =
8$, the dashed one is for $\Gamma = 12$ and $\Gamma = 13$, and the
dots line is for $\Gamma = 16$. The zoom-in panel shows a detail of the same
plot at low values of $\theta$. } \label{theta}
\end{figure}

In Figure \ref{theta} we show the evolution of the fractional polarization
in the observer system with the viewing angle. The angle is corrected for
aberration. In the zoom-in panel we show the detail of the changes for
small viewing angles as expected for 3C279. It can be seen that just a
small change in the orientation of the shock velocity can produce a large
change in the degree of polarization. In order to get a rapid variation of
$\sim10$ \% as we observed, a change of $\sim 2\deg$ is enough. The
dependence on $\Gamma$ is not very strong for small angles. A deviation of
the shock on small timescales can be the effect of an helical magnetic
field produced by the jet precession \citep{Rol94}.

\begin{figure}
\includegraphics[width=\hsize]{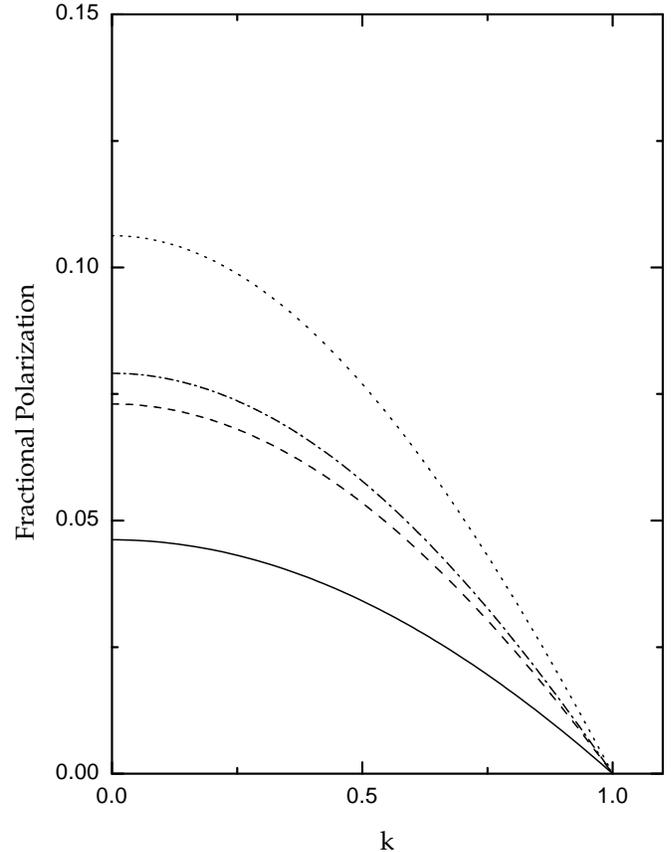}
\caption{Variation of the fractional polarization as a function of
the factor of compression of the plasma. The solid line is for
$\theta^{'}=20^\circ$, the dashed one is for $\theta^{'}=25^\circ$,
the dash-dots line is for $\theta^{'}=26^\circ$, and the dots are
for $\theta^{'}=30^\circ$.} \label{pik}
\end{figure}

Alternatively to the purely kinematic model of \citet{G92}, rapid changes in
the polarization might result from turbulent effects that introduce rapid
changes in the compression factor $k$ \citep{M92}. In Figure \ref{pik} we
show the fractional polarization $\Pi$ as a function of $k$ for different
aberrated viewing angles. From the figure it is clear that even very
strong variations in $k$ cannot produce large changes in
$\Pi$. Hence we favor a kinematic model with a changing viewing angle as the
best explanation for the observed behaviour of the polarization at
internight timescales.  Nonetheless, the rapid intranight flickering can be
the effect of turbulence (and the consequent fluctuation of the compression
ratio) in the post-shock region.

The sudden change in the polarization angle observed between the
nights of March 11 and 12, 2002, might be the result of the injection
of a new shock in the jet. The shock compresses the magnetic field
parallel to the shock front producing a sudden change in the position
angle of the polarized synchrotron flux. The infrared observations
show a decrease of the flux that accompanies the decrease in the
degree of polarization. If a new shock was injected in early March 12,
then the flux should have raised afterwards. Unfortunately, due to the
weather conditions, no observations were possible that night at LNA
Observatory. However, the achromatic 0.4 magnitudes variation in the
near infrared flux 0.4 magnitudes observed in April, 26 and the
chromatic variations in May 18-19, are compatible with short lived
shocks, with relatively high formation rates.

\section{Conclusions \label{s_conc}}

We have carried out polarization observations with very high time
resolution of the OVV blazar 3C279. A large variation of $\sim10$\% in the
degree of linear polarization was observed on internight timescales. More
rapid flickering, with timescales from minutes to hours, was also present
within each single night. Simultaneous infrared observations indicate
that the total synchrotron flux was decreasing while the fractional
polarization was also decreasing. This overall behaviour and
IR observations on larger time scales seems to agree
with what is expected from a relativistic shock that changes mildly the
viewing angle.  The rapid flickering might be due to turbulence in the
post-shock region. New simultaneous multiwavelength and polarization
observations of this extremely active source can shed further light on
the interactions of relativistic shocks with small features or bends in
the inner jet.

\begin{acknowledgements}
Research on AGNs with G.E Romero is mainly supported by
Fundaci\'on Antorchas. Additional support was provided by the
Agencies CONICET (PIP 0430/98) and ANPCT (PICT 98 No. 03-04881).
GER thanks the kind hospitality of the Max-Planck-Institut f\"ur
Kernphysik (Heidelberg) where the final part of this work was
completed. TPD and ZA acknowledges FAPESP support (98/07491-8 and
00/06769-4) and CNPq (306062/88). We thank constructive comments
by Dr. L. Takalo.

\end{acknowledgements}

%\bibliography{QSO}
%\bibliographystyle{aa}

\end{document}